# Some Critical Thinking on EV Battery Reliability: from Enhancement to Optimization

- comprehensive perspectives, lifecycle innovation, system cognation, and strategic insights

Jing (Janet) Lin[1,2*], Christofer Silfvenius[3]

1. Division of Operation and Maintenance, Luleå University of Technology, 97187, Luleå, Sweden
2. Division of Product Realization, Mälardalen University, 63220, Eskilstuna, Sweden
3. SCANIA Technical Center, SCANIA AB, 15132, Södertälje, Sweden

**Abstract:** In the coming era of sustainable transportation, electric vehicles (EVs) are emerging as a pivotal component in the transition towards the eco-friendly mobility paradigm shift. Central to this shift is the evolution of EV battery technology, which not only influences the economic viability of these vehicles but also dictates their performance, efficiency, and environmental impact. This study focuses on EV battery reliability, a vital aspect that significantly determines the sustainability, performance, and longevity of these batteries.

Current efforts to improve EV battery reliability are predominantly targeting specific isolated areas such as design, manufacturing, power density optimization, user behaviour, charging strategies, and environmental impacts. Such approaches, while beneficial in addressing individual challenges, often overlooks the comprehensive interplay within holistic and sustainable systems, with a potential risk of short-term solutions and future long-term issues. This study is structured around three key research questions that aim to address a holistic lifecycle optimisation for EV batteries including improved EV battery reliability.

Through an in-depth exploration across micro, meso, and macro perspectives, this study examines interconnected factors influencing battery reliability, highlighting the need for a broader ecosystem analysis. This study presents an innovative lifecycle framework by focusing on the use phases (use, reuse, repurpose) including recycling but also introducing the concept of "Zero"-Life reliability. Furthermore, the study delves into system cognition, conceptualizing reliability ecosystems and lifecycle frameworks through a dynamic, multi-dimensional approach; more importantly, revealing the intrinsic nature of reliability system optimization.

Ultimately, this study aims to bridge the gap between isolated reliability improvement and a holistic system approach. It offers insights into optimizing the EV battery reliability system, thereby aligning with global sustainability objectives, and advancing the field of sustainable transportation and EV technology.

---

[*] Corresponding Author: Jing (Janet) Lin, Email: janet.lin@mdu.se; janet.lin@ltu.se.

**Chapter 1.  Introduction**

In the transformative era of sustainable transportation, electric vehicles (EVs) stand at the forefront of an eco-conscious e-mobility revolution, which is not just about replacing fossil fuel-driven vehicles with electric ones; it's about fostering a sustainable, efficient, and environmentally friendly transportation system that aligns with broader goals of reducing emissions, conserving energy, and promoting public health and economic growth (European Parliament, 2023; ESG, 2023; McKinsey , 2023; TRATON, 2022; Global Battery Alliance, 2020).

EVs distinguish themselves from traditional vehicles by featuring a novel and crucial electrical system, which include battery packs, charging mechanisms, and power distribution systems. This battery system serves as the primary power source for propulsion.  The battery system is not only economically significant, accounting for up to 40% of the total cost of EV (Thatcher, 2022), but also fundamentally influencing the EV's driving range, performance, efficiency, and environmental impact. Despite technological advancements (TRATON, 2022), including the lithium battery technology, the next challenge lies in making these batteries more sustainable, high-performing, and durable (European Parliament, 2023).

Reliability, fundamentally defined as the ability of a system to perform its required functions under specific conditions for a designated period, plays a pivotal role in making EV batteries more sustainable, high-performing, and durable by ensuring consistent and safe operation over an extended lifespan, which reduces the need for frequent maintenance and eventual replacement. This long-term dependability is crucial for sustaining high performance under various conditions, minimizing environmental impact through less frequent resource extraction and waste, and ultimately supporting the durability of the batteries, making them a more sustainable choice for electric mobility (Shahriar, et al., 2023).

Current state-of-the-art reviews indicate that efforts to enhance EV battery reliability predominantly focus on isolated solutions, including but not limited to: 1) Battery design, encompassing material selection, uniformity and degradation characters among battery cells and packs (Capasso, et al., 2024; Xia, et al., 2019; Shu, et al., 2020; McGovern, et al., 2023; Milojevic, et al., 2021); 2) Power density optimization (Khan, et al., 2023); 3) Environmental impact factors such as temperature, moisture, and road conditions (Tang, et al., 2023; Gao, et al., 2024; Cheng, et al., 2022); 4) Driving behaviours, including driving schedule, speed, acceleration, and braking patterns (Chou, et al., 2023); 5) Charging and discharging strategies (Lee & Kang, 2023; Zhang, et al., 2024; Yang, et al., 2023; Ma, et al., 2023; Anand, et al., 2020). In these research studies, efficient battery management system (BMS) for condition monitoring, thermal management, balance management, fault diagnosis and prediction (Xiong, et al., 2020; Youssef, et al., 2023; Ali, 2023; Liu, et al., 2022; Yifan, et al., 2024; Waseem, et al., 2023), and metrics such as Battery State-of-Health (SOH), State-of-Charge (SOC), State-of-Energy (SOE), and Remaining Useful Life (RUL) are commonly utilized to gauge operational reliability[1] (Marghichi, et al., 2023; Yi, et al., 2023; Chen, et al., 2023; Lv, et al., 2024; Zhao, et al., 2023).

The reliance on isolated solutions for enhancing EV battery reliability, though beneficial for targeted issues, can fall short in a broader context. This approach often has a limited scope, addressing specific components or aspects without considering the intricate interdependencies within the battery and vehicle systems. Such a narrow focus can lead to *short-term* solutions that fail to tackle underlying systemic challenges, potentially leading to recurring problems. For instance, improvements in battery materials for increased capacity might adversely affect thermal management, highlighting how changes in one area can have unintended consequences in others. Another significant limitation of isolated approaches is their inability to keep pace with rapid advancements in technology and evolving user needs, reducing their *long-term* applicability. Concentrating solely on reliability metrics like State-of-Health (SOH) or Remaining Useful Life (RUL) often overlooks crucial factors such as safety, environmental impact, and user experience. For example, a battery with a high SOH might still pose safety risks if not designed with comprehensive safety mechanisms. The lack of a holistic perspective, incorporating the reliability ecosystem and lifecycle framework, results in missed opportunities for systematic improvement and a comprehensive understanding of the battery's operational environment.

---

[1] Operational reliability = inherent reliability + use reliability. Reliability is a birth to death concern, where: inherent reliability of product is decided in the stages of concept & definition, design & development, manufacture & install; while use reliability of product is decided in the stage of operation & maintenance. For more details, see Chapter 2.

To overcome the above shortcomings on continuous enhancement of EV battery reliability and further propose a holistic approach, this study is structured to answer three research questions (RQs):

- **RQ1: Why is enhancing EV battery reliability a must from a comprehensive and holistic perspective?**
    - How are EV battery reliability ecosystems defined at micro (individual), meso (industrial), and macro (society) levels?
    - Which factors interconnect and influence battery reliability at micro, meso and macro levels?
- **RQ2: What is the framework of EV battery reliability lifecycle?**
    - What is the difference between general reliability framework and lifecycle of EV batteries during various stages?
    - How to understand the difference between "Zero"-Life reliability and reliability?
    - How is the concept of Use and Reuse, Repurpose, and Recycle connecting to the use reliability of EV batteries?
- **RQ3: What is a holistic system cognition of EV battery reliability for reliability system optimization?**
    - What is the composition of system cognation from point to System of Systems (SoS)?
    - How to understand the intrinsic nature of reliability system optimization?

To answer RQ1, Chapter 2 delves into the multifaceted necessity of high EV battery reliability, exploring its importance from micro, meso, and macro perspectives. It provides ecosystem analysis, illustrating how factors interconnect and influence battery reliability. The goal is to underscore the far-reaching impact of reliability on everything from individual consumer satisfaction, industrial competitiveness, to broader environmental sustainability. To answer RQ2, Chapter 3 introduces an innovative approach to understanding the EV battery reliability lifecycle. It transitions from a general reliability framework to a specific focus on the lifecycle of EV batteries, including integrating focuses during the use stages (use, reuse, repurpose, recycle), and a novel concept of "Zero"-Life reliability is introduced—a stage that encompasses the period between battery manufacture to its initial operational use in a vehicle from its intended first lift within the lifecycle framework. To answer RQ3, Chapter 4 presents an integrated view of the reliability ecosystem and lifecycle frameworks of EV batteries, conceptualized through triangular geometries. It explores the concept of time as a pivotal factor that transforms the stationary status into a dynamic one, revealing the intrinsic nature of reliability system optimization. Following the above critical thinking, Chapter 5 engages in an extensive discussion on future directions, offering strategic insights into enhancing system reliability through continuously improvement of reliability system. Through this work, it is aimed to contribute novel insights and practical strategies to the ongoing discourse in sustainable transportation and EV technology, providing a pathway for enhancing EV battery reliability in alignment with global sustainability objectives.

## Chapter 2. The Crucial Quest for High Reliability in EV Batteries: Comprehensive Ecosystem Analysis using Micro, Meso, and Macro Perspectives

This chapter begins by introducing the concept of the "Reliability Ecosystem" in the context of EV batteries, laying the foundation for the subsequent analysis. It then systematically unpacks this ecosystem using micro, meso, and macro dimensions (Fig 2.1). Through this micro to macro approach, the chapter aims to provide insights into why improved EV battery reliability is necessary from comprehensive perspectives (RQ2) through a thorough understanding of "Reliability Ecosystems" which is vital in the pursuit for high reliability in EV batteries (Table 2.1). This multi-perspective analysis not only highlights the complexities involved in achieving high reliability but also underscores the collaborative effort required between different stakeholders to drive the EV industry towards a more reliable, efficient, and sustainable future.

### 2.1 Reliability Ecosystem of EV Batteries

The "Reliability Ecosystem" of EV batteries extends beyond the physical battery unit. The reliability ecosystem refers to a comprehensive network of factors, stakeholders, processes, and interactions that collectively contribute to and influence the final reliability of EV batteries. This ecosystem encompasses multiple layers and dimensions, each playing a crucial role in determining how reliable the batteries are throughout their lifecycle (Fig. 2.1).

- The micro perspective focuses on the individual consumer experience at a single vehicle level, forming the foundation of the individual experience and emphasizing the direct impact of reliability on the end user.
- The meso perspective widens the lens to industry stakeholders, including manufacturers, suppliers, and service networks, considering industry and market dynamics, and showcasing how decisions and innovations at this level directly influence battery quality and reliability.
- Finally, the macro perspective broadens the scope further to encompass societal, environmental, and economic considerations, highlighting the role of policies, global market trends, and environmental regulations in shaping battery reliability.

The interplay between the micro, meso, and macro perspectives highlights the dynamics of improving EV battery reliability. Understanding the importance of EV battery reliability through this tripartite ecosystem provides a comprehensive view of the multiple and significant facets and levels.

- At the micro level, individual consumer experiences and feedback shape manufacturers' understanding of real-world performance and reliability needs. This communication informs meso-level decisions. Consumer satisfaction and market demand at this level also play a crucial role in steering policy decisions and market strategies at both meso and macro levels, reflecting a feedback loop where individual experiences influence broader market trends and acceptance.
- Decisions and innovations at the meso level, including manufacturers, suppliers, and service networks, directly impact the quality and reliability of batteries experienced at the micro level. The industry's approach to enhancing battery reliability, motivated by competition and regulatory adherence, extends its influence to broader economic and environmental impacts. This includes setting industry standards and influencing policy decisions that resonate at the macro level.
- At the macro level, policies, environmental regulations, and economic incentives guide both consumer behaviour (micro) and industry practices (meso). This includes initiatives promoting sustainability and emissions reductions, which can drive the demand for more reliable batteries. The macro perspective encompasses the cumulative impact of actions taken at both the micro and meso levels, considering how enhancing battery reliability contributes to global sustainability goals, economic growth in green technologies, and societal well-being.

These interconnected perspectives defines a comprehensive ecosystem where consumer experiences shape industry practices and standards, industry advancements influence market landscapes and policy environments, and overarching policies and regulations guide consumer and industry behaviours towards sustainable outcomes.

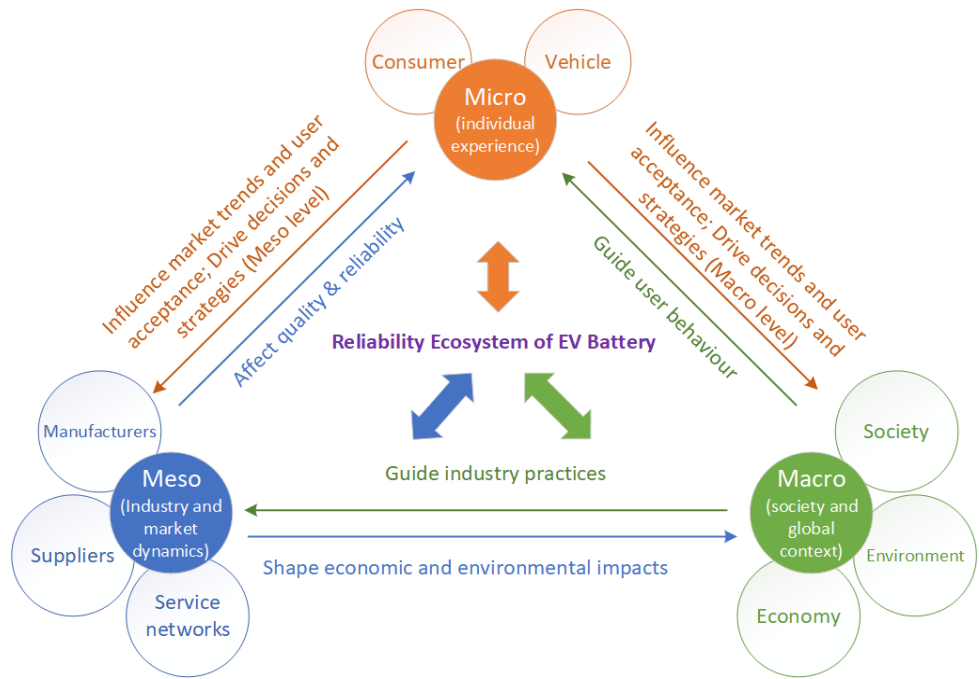

Fig 2.1 The Reliability Ecosystem of EV Batteries

## 2.2 Maximizing Consumer Benefits: The Micro-Level Impact of Enhanced EV Battery Reliability

In the rapidly evolving domain of EV technology, the importance of battery reliability from a micro-level perspective on individual consumers and their interaction with EVs, is paramount. A more granular viewpoint sheds light on how reliability profoundly influences various aspects of the EV experience, encompassing functionality, safety, comfort, cost-effectiveness, and sustainability.

- **Functionality and Performance:** High reliability in EV batteries is essential for ensuring consistent and predictable vehicle performance. This reliability is fundamental to providing consumers with a dependable range per charge, which is crucial for effective planning and execution of journeys without the worry of unexpected power issues. Additionally, the sustained operational efficiency of reliable batteries means that the vehicle's performance remains robust over time, enhancing the overall driving experience.
- **Safety Assurance:** Safety is intrinsically linked to the reliability of EV batteries. High reliability drastically reduces risks such as on-road power failures or battery-related fires, safeguarding not only the vehicle occupants but also ensuring overall road safety. As EVs continue to increase in numbers, the reliability of their batteries becomes a cornerstone in building and maintaining public trust in this emerging technology.
- **Comfort and User Experience:** Reliable batteries significantly enhance the driving pleasure by mitigating range anxiety and ensuring consistent performance of vehicle features, thereby elevating overall user satisfaction. This reliability also provides consumers with a peace of mind, knowing their vehicle is dependable and will perform as expected, a crucial factor in the widespread adoption of EVs.
- **Economic Considerations:** From an economic standpoint, the battery system, often being the costliest component of an EV, plays a significant role in determining the total cost of ownership. High reliability translates into fewer needs for battery replacements and lower maintenance costs throughout the vehicle's lifespan, offering substantial economic benefits. Furthermore, EVs known for their reliable batteries tend to maintain higher resale values, thereby providing an additional economic advantage to consumers.
- **Sustainability Impact:** Reliability in EV batteries align with the growing consumer awareness and demand for environmental sustainability. By reducing the frequency of battery replacements, reliable batteries diminish the environmental burden associated with their production and disposal. Additionally, these batteries contribute to sustainable mobility by enhancing the longevity and performance of EVs, in line with global efforts to reduce carbon emissions and foster sustainable transportation.

This chapter underscores that the reliability of EV batteries at the micro level transcends being merely a technical attribute; it is a pivotal factor that significantly impacts the consumer experience, vehicle safety, economic value, and environmental sustainability. The advancements in battery reliability manifest in tangible benefits for consumers, promoting increased adoption of EVs and contributing to a more sustainable and user-centric future in mobility.

## 2.3 Industry Evolution: The Meso-Level Influence of Advancing EV Battery Reliability

The meso-level landscape within the EV battery reliability ecosystem is characterized by the integral roles and responsibilities of industry stakeholders, including manufacturers, suppliers, and service networks. This section of the chapter delves into the profound significance of high reliability in EV batteries from a meso-level viewpoint, highlighting its substantial impact on industry dynamics, supply chain management, and the broader market.

- **Enhancing Brand Reputation and Competitive Advantage:** At the heart of industry evolution lies the enhancement of brand reputation and competitive advantage. For manufacturers, the reliability of their EV batteries is deeply entwined with their brand reputation. High reliability is equated with quality and plays a crucial role in fostering consumer trust, a fundamental aspect of building brand loyalty and establishing a strong market presence. Additionally, in the rapidly evolving EV market, the reliability of batteries emerges as a critical factor for competitive differentiation. Manufacturers committed to consistently delivering highly reliable batteries position themselves at a competitive advantage, attracting a larger customer base and securing a more dominant market position.
- **Economic Implications: Warranty Costs and Efficiency:** The economic implications of high battery reliability are significant, especially in terms of warranty costs and operational efficiency. Enhanced reliability directly leads to a decrease in warranty claims, as lower rates of battery failures and recalls translate into reduced after-sales service expenses. This reliability further influences operational efficiency, where reliable production processes and products minimize the need for frequent quality checks and rework, thereby streamlining operations and yielding cost savings.
- **Impact on the Supply Chain:** The impact of high battery reliability on the supply chain is also profound. It necessitates stringent quality standards across the entire supply chain, encompassing rigorous testing of components and materials to ensure that every element of the battery adheres to high reliability criteria. This focus on reliability extends to logistics and distribution strategies, where efficient handling and transportation methods are crucial to maintain the integrity of batteries from production to consumer use.

- **Influencing Industry Standards and Regulations:** Advancements in battery reliability at the meso-level play a vital role in influencing both industry standards and regulations. Manufacturers leading in reliability often establish benchmarks that gradually evolve into industry standards, guiding practices, and product development across the sector. High reliability also aligns with, and frequently surpasses, regulatory standards. As regulations around EV batteries become more stringent, manufacturers prioritizing reliability find themselves well-positioned in terms of compliance, thus avoiding potential legal and financial repercussions.
- **The Ripple Effect to Micro and Macro Levels:** This meso-level commitment to reliability also has a ripple effect that extends to both micro and macro levels. At the micro-level, advancements in battery reliability enhance consumer experiences, influencing individual preferences and behaviours. At the macro-level, the industry's approach to reliability contributes to shaping broader economic and environmental trends, influencing market dynamics, guiding policy decisions, and shaping the regulatory landscape. This ultimately affects the global transition towards sustainable transportation.

This chapter emphasizes the critical importance of the meso-level in advancing EV battery reliability. It highlights how industry players' dedication to reliability not only influences their immediate operational environment but also has far-reaching implications across the broader market and regulatory frameworks.

**2.4 Strategic Sustainability: The Macro-Level Advantages of Boosting EV Battery Reliability**

In the expansive macro-level landscape of the EV battery reliability ecosystem, this chapter explores the overarching significance of reliability within the context of sustainability. It delves into the intersection of enhanced EV battery reliability with macro-level sustainability objectives, economic strategies, and commitments to environmental stewardship.

- **Conservation of Finite Resources:** The conservation of finite resources is a critical aspect at macro level. The finite nature of essential raw materials like lithium and cobalt necessitates the extension of battery life. High reliability in EV batteries plays a pivotal role in prolonging their usable lifespan, thus conserving scarce resources, and reducing the environmental footprint associated with extraction and processing.
- **Reduce Environmental Impacts:** Addressing environmental impacts forms another key facet of this analysis. Reliable batteries are instrumental in reducing the carbon footprint of electric vehicles by ensuring longer life cycles and reducing the frequency of manufacturing, thereby curtailing emissions related to production. Moreover, by extending the operational lifespan of EV batteries, the frequency of battery disposal is significantly lowered, reducing the generation of e-waste, and mitigating its environmental impact.
- **Reduce Pressure on Recycling Systems:** The role of high reliability in alleviating pressure on recycling systems is also examined. Enhanced reliability diminishes the strain on recycling infrastructures by lowering the rate of battery turnover. This not only promotes economic efficiency through reduced manufacturing and end-of-life handling costs but also contributes to environmental conservation. Furthermore, the potential for repurposing batteries for applications such as energy storage post-vehicle use exemplifies the principles of the circular economy, maximizing the utility of these materials and reducing waste.

- **Influencing Global Policies and Market Dynamics:** The influence of battery reliability on global policies and market dynamics is profound. The pursuit of reliable EV batteries aligns with international sustainability and clean energy initiatives, shaping policy decisions and guiding regulations towards more sustainable practices. This push for reliability also drives market trends, steering the automotive industry toward environmentally responsible solutions.

This chapter broadens the scope of impact to consider the implications beyond individual benefits. High reliability in EV batteries emerges as a strategic imperative for global environmental sustainability, transcending the benefits realized at micro and meso levels. The discussion in this chapter emphasizes the integral role of EV battery reliability within the broader context of sustainable development and environmental conservation, highlighting how advancements in this area are crucial not only for immediate stakeholders but for the overarching goals of sustainable progress.

Table 2.1 Crucial Quest for High Reliability in EV Batteries

| Levels | Impacts | Sub contents |
|---|---|---|
| Micro-Level | Functionality and Performance | Consistent and Predictable Performance |
| | | Long-Term Operational Efficiency |
| | Safety Assurance | Mitigation of Safety Hazards |
| | | Trust in Technology |
| | Comfort and User Experience | Enhanced Driving Pleasure |
| | | Peace of Mind |
| | Economic Considerations | Cost-Effectiveness |
| | | Resale Value |
| | Sustainability Impact | Alignment with Environmental Values |
| | | Contribution to Sustainable Mobility |
| Meso-Level | Enhancing Brand Reputation and Competitive Advantage | Brand Image and Consumer Trust |
| | | Competitive Differentiation |
| | Economic Implications | Reducing Warranty Claims |
| | | Operational Efficiency |
| | Impact on the Supply Chain | Quality Standards for Suppliers |
| | | Logistics and Distribution |
| | Influencing Industry Standards and Regulations | Setting Industry Benchmarks |
| | | Regulatory Compliance |
| | The Ripple Effect to Micro and Macro Levels | Feedback to Consumer Level |
| | | Shaping Market and Policy Trends |
| Macro-Level | Conservation of Finite Resources | Importance of Longevity |
| | | Impact on Resource Extraction |
| | Reducing Environmental Impacts | Mitigating Emissions |
| | | Minimizing Waste |
| | Alleviating Pressure on Recycling Systems | Economic and Environmental Efficiency |
| | | Facilitating Circular Economy |
| | Influencing Global Policies and Market Dynamics | Shaping Sustainable Policies: |
| | | Driving Market Trends |

# Chapter 3: Unveiling Operational Reliability: Transitioning Lifecycle Frameworks from General Asset to EV Battery

After introducing the general asset lifecycle framework, this chapter provides a detailed exploration of the EV battery reliability lifecycle (RQ2), which enrich our understanding of both inherent reliability and use reliability for EV batteries.

This chapter also introduces and elaborates on the concept of "Zero"-Life Reliability as well as focusing on the use stages (use, reuse, repurpose, and recycle) and integrates this into an innovative lifecycle framework. This pioneering approach not only enhances the reliability of EV batteries from the outset but also contributes significantly to the broader ecosystem, influencing future innovations, operational strategies, and sustainability efforts in the realm of electric mobility.

## 3.1 Operational Reliability: Understanding General Asset Lifecycle Framework

The operational reliability of a system (or component) during its operational phase, also known as overall reliability, is determined by a combination of factors from both its design and manufacturing stage (inherent reliability) and its actual usage conditions (use reliability). Operational reliability is the combination of inherent reliability and use reliability, capturing the complete spectrum of a product's performance.

Achieving high operational reliability demands not only a focus on outstanding design and manufacturing for inherent reliability but also a comprehensive understanding and adaptation to real-world usage to ensure long-term use reliability. As such, operational reliability provides a more accurate measure of a product's overall reliability, considering the multitude of variables and conditions that influence performance outside of controlled environments.

The general reliability for a system or component, often described as a "birth to death" framework, follows a product from its initial conception through to its eventual failure and disposal (Fig 3.1). This framework is commonly used in various industries to manage and understand the reliability of physical assets. A breakdown of the key stages is shown in Table 3.1. In the general lifecycle framework, reliability is a concern at every stage, from ensuring a robust design that can withstand operational demands to maintaining the asset effectively during its use phase. The goal is to maximize the asset's useful life while minimizing downtime and failures.

It should be noticed, having prepared the best possible engineering design, we can still be limited by the inherent reliability, if we ignore the necessity of stringent quality control during the manufacture. Quality in its simplest interpretation is reliability during manufacturing phase and ensures that only proper materials, processes, and quality control techniques have been used.

Table 3.1 Breakdown of the key stages of general asset reliability lifecycle framework

| Stage | | Name | Contents |
|---|---|---|---|
| Overall/Operational Reliability | Inherent Reliability | Concept & Definition (Birth) | This stage involves defining the asset's purpose, specifications, and design. Decisions made here have a long-term impact on the asset's reliability. |
| | | Design & and Development | Engineers and designers work to create a product that meets specified requirements, including reliability targets. This involves selecting materials, determining the architecture, and considering manufacturing processes. |
| | | Manufacture & Production | **Manufacturing Process:** The asset is produced or constructed. Quality control and assurance are critical at this stage to ensure that the design specifications and reliability standards are met.<br>**Testing and Validation:** Before the asset is released, it undergoes rigorous testing to identify any defects or reliability issues that need to be addressed. |
| | | Distribution | The asset is transported to its final location or to consumers. |
| | | Installation & Commissioning | Proper **installation** is crucial for ensuring the asset functions as intended. **Commissioning** involves checking that all systems operate correctly. |
| | Use Reliability | Operation & Maintenance | **Active Use:** The asset is put into service and begins its operational life. This phase can last for years or decades, depending on the asset.<br>**Maintenance & Repairs:** Regular maintenance is essential for preserving reliability. Unplanned repairs may occur due to unforeseen failures. |
| / | / | End of Life & Disposal (Death) | **Decommissioning:** The asset is taken out of service. This may occur due to obsolescence, failure, or because it's no longer economically viable to maintain.<br>**Disposal/Recycling:** The asset is dismantled, and materials are disposed of or recycled. Environmental and safety considerations are paramount in this stage |

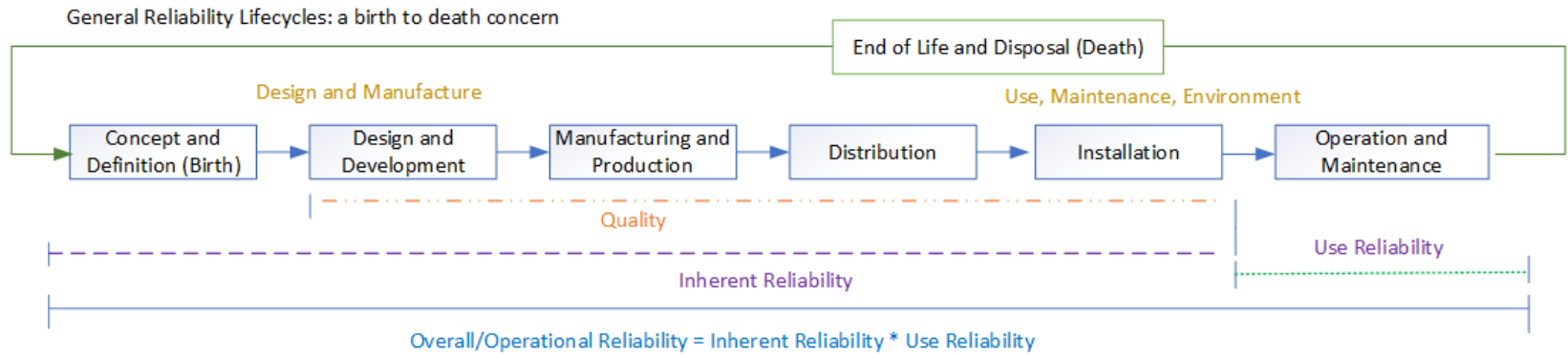

Fig 3.1 General Asset Reliability Lifecycle Framework

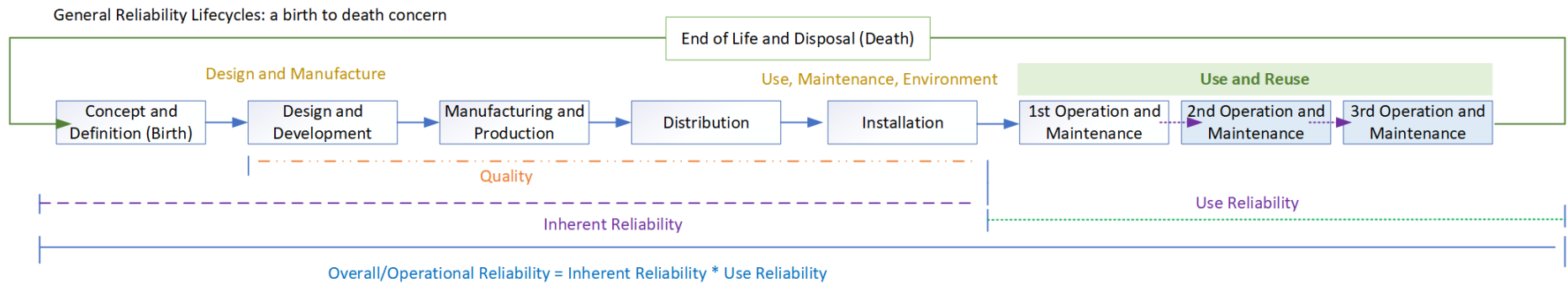

Fig 3.2 EV Battery Reliability Lifecycle Framework with 1st, 2nd, and 3rd life added

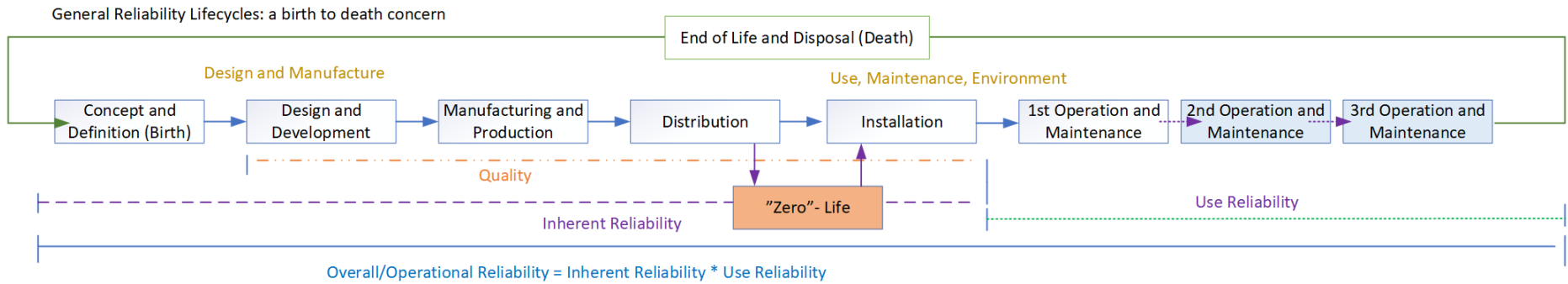

Fig 3.3 EV Battery Reliability Lifecycle Framework with Zero$^{th}$, 1$^{st}$, 2$^{nd}$, and 3$^{rd}$ life added

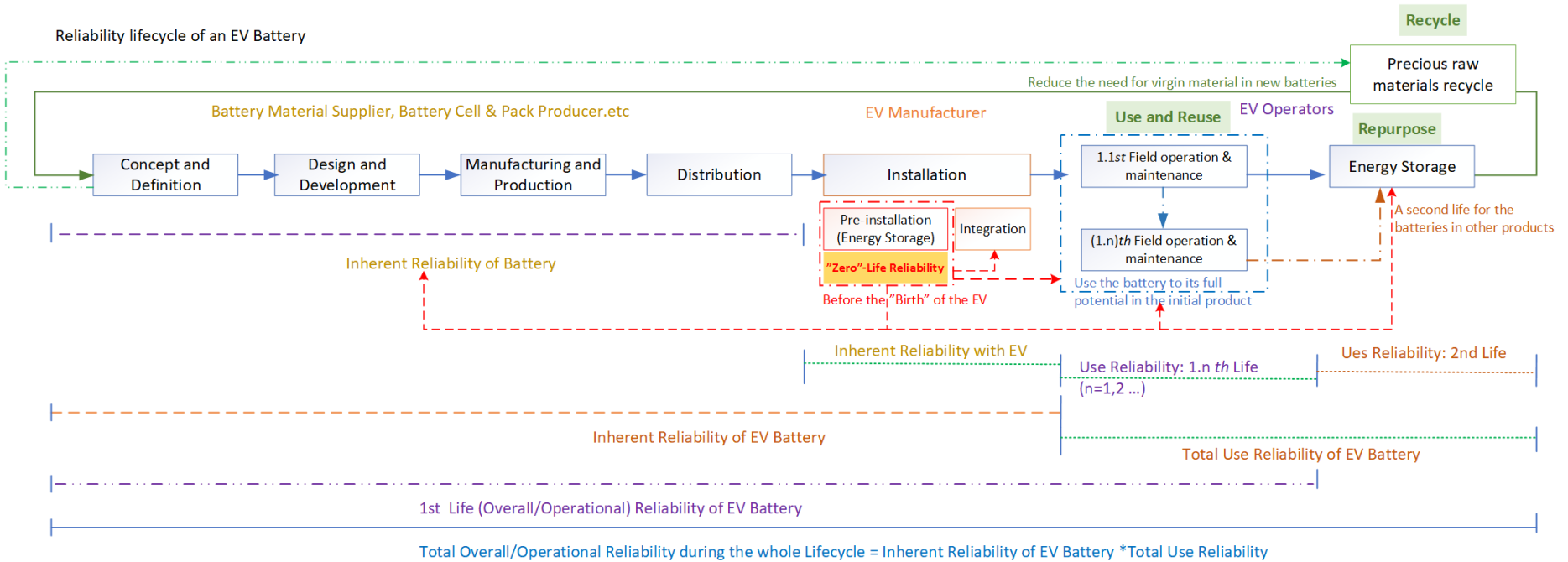

Fig 3.4 EV Battery Reliability Lifecycle Framework including more details of a "Zero"-Life stage

## 3.2 EV Battery Reliability Lifecycle Framework: novel insights on operational reliability

The lifecycle of an electric vehicle battery presents a distinctive framework for understanding the nuances of overall reliability, a perspective that differs considerably from the general asset lifecycle framework. As illustrated in Fig 3.2 to 3.4, the EV battery lifecycle offers unique insights into both inherent reliability and use reliability.

From inherent reliability perspectives, the concept of inherent reliability in the context of EV batteries encompasses two key components. Firstly, there is the inherent reliability of the battery itself, which includes both cells, modules, climatization, control system and the battery pack. This aspect focuses on the reliability engineered into the battery during its design and manufacturing stages. Secondly, inherent reliability is also linked to the integration of the battery within the EV by the manufacturer, highlighting the interdependence between the battery and the vehicle's overall design and functionality.

### "Zero"- Life

The concept of "Zero"-Life reliability in EV batteries, introduced in Fig 3.3, represents a groundbreaking and essential stage in the assessment of battery inherent reliability. This concept refers to the period from the battery's manufacture up to the point where it is installed and begins operation in an electric vehicle. This study proposes a novel and cost-effective approach, for vehicle manufacturers to begin utilizing EV batteries for energy storage purposes during the critical pre-installation stage, rather than simply storing them as inventory. This practice would not only serve as an efficient utilization of resources but also starts the operational lifecycle of the battery earlier, offering cost savings in terms of reduced tests for manufacturers. This innovative stage plays a pivotal role in establishing the foundational reliability of EV batteries. A detailed exploration of this novel concept is provided in section 3.3, where its significance and impact on the overall reliability of EV batteries are thoroughly examined.

From use reliability perspectives, the lifecycle of an EV battery encompasses several stages, each contributing uniquely to its use reliability. Notably, there is the "use and reuse" stage (Fig 3.2-3.4), which sees the battery achieving multiple lifecycles within its initial product application. For instance, in Fig 3.4, a battery may have multiple lifetimes when installed in different EVs or other consequents fewer demanding applications. Since they continue to be utilized in vehicles, these various lifetimes are referred to as 1. $n$ ($n \geq 1$) use life. Additionally, there is the "repurpose" stage, commonly seen in the form of energy storage applications, effectively constituting a second life for these batteries in different products. Consequently, use reliability can be conceptualized as comprising the 1. $n$ (where $n = 1,2, ...$) use reliabilities (encompassing multiple life cycles within EVs).

While the disposal stage often receives limited attention in general reliability lifecycles, it holds a critical role in the EV battery reliability lifecycle. The "Recycle" stage, where raw materials from spent batteries are reclaimed for new batteries, provides pivotal reliability information, particularly under the lens of the sustainability development goals (SDGs). This stage underscores the growing importance of sustainable practices in the lifecycle of EV batteries.

The interconnections among inherent reliability, total use reliability, and recycling in the context of EV batteries form a continuous loop that reflects the entire lifecycle of the battery, from production to end-of-life (Fig 3.4). Here's how they are interconnected:

- **Inherent Reliability as the Foundation:** Inherent reliability is established during the design and manufacturing stages of the EV battery. It sets the baseline for how the battery will perform under both normal and challenging conditions. This initial reliability is crucial as it determines the robustness, efficiency, and durability of the battery right from the start. High inherent reliability can lead to longer battery life, fewer early-life failures, and potentially more effective recycling outcomes.

- **Total Use Reliability as the Operational Extension:** Total use reliability encompasses the battery's performance during its entire number of operational lives, including initial use in an EV and any subsequent reuse or repurposing, such as energy storage. The inherent reliability of the battery influences its total use reliability; a battery with high inherent reliability is more likely to maintain good performance over its lifecycle. This stage is where the battery experiences real-world conditions, and its performance can be affected by various factors like usage patterns, maintenance, environmental conditions, and ageing.
- **Recycling as the End-of-Life Process:** The end-of-life stage of the battery, particularly recycling, is influenced by both the inherent reliability and the total use reliability. A battery with high inherent reliability may retain more of its original capacity and quality, making it a more valuable resource for recycling. The condition of the battery at the end of its total use reliability stage (after primary and secondary uses) will determine the ease and efficiency of material recovery during recycling.

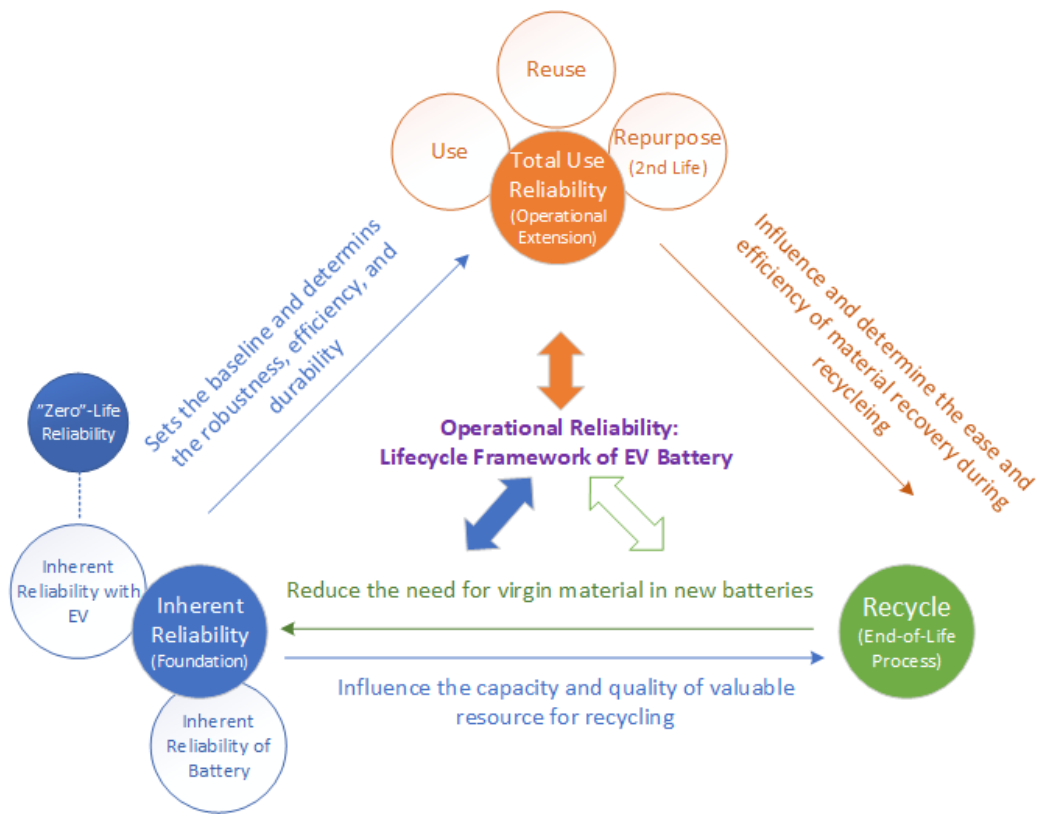

Fig 3.5 The Operational Reliability: Lifecycle Framework of EV Batteries

In summary, inherent reliability impacts total use reliability, which in turn influences the efficacy and value of recycling. This interconnection underscores the importance of considering the entire lifecycle when designing and manufacturing EV batteries, aiming not only for immediate performance but also for long-term sustainability and recyclability. Each stage is interdependent, with early decisions impacting the later stages, thereby creating a cycle that is crucial for sustainable battery management and environmental stewardship in the EV industry.

A comprehensive breakdown of key studies conducted during these various stages is presented in Tables from Table 3.2.1 to Table 3.2.7. This table offers an in-depth look into how each stage of the EV battery lifecycle contributes to our understanding of operational reliability, highlighting the unique challenges and opportunities that arise at each phase. For clarity, it should be noted that Tables 3.2.1 to 3.2.7 do not specifically represent the micro, meso, or macro perspectives.

### 3.3 "Zero"-Life Reliability of EV Battery: a Unique Focus from Inherent Reliability

As mentioned before, innovative manufacturing strategies can lead to the utilization of EV batteries for energy storage during their critical pre-installation phase, a shift from traditional inventory storage practices. This strategy not only initiates the operational lifecycle of batteries earlier but also offers significant cost benefits. By engaging batteries in active use before installation in vehicles, manufacturers enhance resource efficiency in the production process.

During this period, which can last up to twelve months to include seasonal variations, the batteries undergo thousands of charge and discharge cycles, a process that serves triple purposes. Besides the above energy storage, this extensive usage allows for the collection of comprehensive data, crucial for quality assurance. This data ensures that each battery not only meets the highest standards of quality before being integrated into an EV; moreover, and perhaps more significantly, the data gathered from these cycles facilitates a transition from traditional statistical process control (SPC) methods to a more forward-looking statistical process prediction (SPP) approach. Last but not the least, this shift enables manufacturers to perform detailed reliability analyses, predicting the battery's performance and lifespan across various stages of its use, including the 1.n (multiple use cycles within the EV) and 2nd lifetimes (post-EV usage in other applications).

Correspondingly, in the evolving narrative of EV battery reliability, a novel and critical concept emerges (Fig 3.2 - 3.4): the "Zero"-Life Reliability. This concept, first introduced in section 3.2, represents a pivotal stage in the EV battery lifecycle that has not been studied before. It is akin to a missing piece of the puzzle in understanding battery reliability, spotlighting the period from battery manufacture to its initial deployment in the field.

"Zero"-Life Reliability of an EV battery represents a paradigm shift in battery manufacturing and reliability assessment. By effectively utilizing the batteries prior to their installation in EVs, manufacturers are not only optimizing resource utilization but also enhancing their predictive capabilities regarding the battery's future performance and reliability. This approach aligns with the broader goals of sustainability and efficiency in the EV industry, setting a new standard for how battery lifecycles are managed and analysed. Anticipated Outcomes of "Zero"-Life Reliability Analytics include (but not limited to):

- **Quality Assessment Before Installation:** The primary goal is to ensure that EV batteries meet stringent performance and safety benchmarks even before they are installed. This proactive quality assessment aims to pre-emptively address any potential issues and therefore make it possible from SPC to SPP.
- **Prediction of Remaining Useful Lifetime (RUL):** By analysing data from "Zero"-Life stages, it becomes possible to accurately forecast the RUL of EV batteries, extending from their first lifecycle to subsequent ones. This prediction aids in planning for maintenance, replacements, and recycling.
- **Guidance for In-Field Operations and Maintenance:** The insights gained from "Zero"-Life Reliability Analytics provide essential guidelines for optimizing the performance, reliability, and longevity of EVs once they are operational.

- **Feedback Loop for Continuous Improvement:** Establishing a robust feedback mechanism between vehicle manufacturers and battery producers is critical for continuous enhancements in battery design, manufacturing, and quality control.
- **Enriching the Battery Ecosystem:** The data and knowledge gathered during this stage play a vital role in fostering a transparent and enriched battery ecosystem, beneficial for manufacturers, consumers, and researchers.

## 3.4 Total Use Reliability of EV Battery: Redefining Assessment, Prediction, and Improvement

As introduced in Fig 3.2 - 3.4, a significant evolution of EV battery is observed in the concept of reliability, extending beyond the traditional one-use lifecycle. The total use reliability of EV batteries encompasses the stages of use and reuse, repurposing, and recycling, each contributing to a broader, more comprehensive understanding of reliability. This expanded perspective challenges and redefines the conventional methods of reliability assessment, prediction, and improvement.

Traditionally, battery reliability was assessed and predicted based on a single-use phase, typically focusing on the battery's performance from installation to its end-of-life in a vehicle. However, the concept of reuse broadens this assessment. Reuse involves utilizing the battery to its full potential in its initial application, ensuring that the battery's life is maximized in its primary role. This stage demands a reassessment of the battery's performance and reliability as it continues to serve in the same application but with reduced efficacy due to aging and wear.

Once the battery reaches the end of its viable life in a vehicle, it can still hold significant value (Union of Concerned Scienstists, 2020; National Renewable Energy Laboratory (NREL), 2020). The stage of repurposing involves giving the battery a "second life" in different applications, such as in Battery Energy Storage Systems (BESS). These systems, akin to large-scale power banks, can provide services like frequency balancing for grid operators or boosting local electric grids. This phase of the battery's life adds a new dimension to reliability assessment, as the battery now functions in a different context, with varying demands and stressors compared to its initial use in a vehicle (Camboim, et al., 2024).

The importance of reliability assessment for the reuse and repurpose of batteries is crucial for ensuring safety, maximizing utility, maintaining value, and ensuring regulatory compliance in the reuse of batteries, thereby playing a vital role in the sustainable lifecycle management of battery technology.

- **Determines Usability and Lifespan:** Reliability assessment helps determine RUL of the battery. It is critical in ascertaining how long the battery can effectively be used in its new application without significant performance degradation.
- **Ensures Safety:** Assessing the reliability of batteries prior to reuse can help to identify potential risks, such as the likelihood of overheating or fire, which are critical considerations for both the new users and regulatory compliance.
- **Influences Residual Value:** The reliability assessment provides an indication of the battery's current state and potential for future use. This directly impacts the residual value of the battery in the second-hand market.
- **Guides Appropriate Application Selection:** By understanding the reliability and capacity of the battery, decision-makers can better match the battery to suitable reuse and repurpose applications where it can perform effectively, such as in less demanding stationary storage roles.
- **Helps in Warranty and Insurance Decisions:** Reliable data on the battery's health and expected lifespan can guide warranty decisions and insurance policies for reuse and repurpose batteries, impacting both seller's and buyer's confidence.

- **Facilitates Regulatory Compliance:** Many regions have stringent regulations for battery reuse and repurpose. A thorough reliability assessment ensures that the repurposed batteries meet the necessary safety and performance standards set by regulatory bodies.
- **Supports Efficient Resource Utilization:** Understanding the remaining reliability of batteries enables more efficient use of resources. It helps in avoiding premature recycling of batteries that still have significant usable life, thereby contributing to sustainability.
- **Boosts Consumer Confidence:** For consumers, knowing that a battery has been reliably assessed and is still in good condition can be a key factor in their decision to choose products that incorporate reused batteries.

The final stage in the total use reliability framework is recycling. When a battery can no longer be effectively used or repurposed, recycling becomes crucial. This process involves extracting valuable materials like cobalt for use in new batteries, thereby reducing the need for virgin materials. Reliability in this context shifts focus to the efficiency and efficacy of the recycling process and the quality of the recovered materials.

The expansion of reliability assessment to include reuse, repurposing, and recycling stages necessitates new approaches and methodologies. It's no longer sufficient to predict a battery's performance based solely on its first life in a vehicle. Instead, a more holistic view is required, considering the various potential phases of a battery's life. This approach also demands improvements and innovations in battery technology that consider these extended stages of use, focusing on factors like longevity, degradation rate, and adaptability to different uses.

The total use reliability of EV batteries demands a shift from a linear view of battery life to a cyclical one, where each stage of the battery's life is considered integral to its overall reliability. This comprehensive approach not only enhances the sustainability and efficiency of battery use but also aligns with the broader goals of circular economy and environmental stewardship in the EV industry.

Table 3.2.1 Breakdown of the key studies of EV battery reliability lifecycle framework: Concept and Definition

| Character | Inherent Reliability | Stage | Concept and Definition |
|---|---|---|---|
| Highlights | This stage encapsulates the initial ideation, design choices, and planning that set the trajectory for the battery's development. ||||
| Contents | Current research examples ||||
| Initial Ideation & Purpose Definition | Studies here involve analysing market trends, consumer demands, and the forecasted needs of the EV industry to determine the purpose and target specifications for new battery designs.<br>Research is conducted to assess the feasibility of various battery chemistries and configurations, considering factors such as energy density, power output, safety, and cost. ||||
| Design Parameters & Specifications | This study area focuses on optimizing the design for performance, including maximizing energy density, minimizing weight, and ensuring safety. It often involves computational modelling and simulation to predict how design choices will impact reliability.<br>Studies investigate the properties of various materials for battery components (e.g., cathode, anode, electrolyte) to improve performance, longevity, and safety while also considering cost and availability.<br>Research is also devoted to incorporating sustainability into the design phase, such as selecting materials that are more abundant or have less environmental impact during extraction and processing. ||||
| Reliability Targeting | At this stage, engineers focus on incorporating reliability into the design by employing reliability engineering principles. This includes failure mode and effects analysis (FMEA), fault tree analysis (FTA), reliability prediction models, and robust design techniques.<br>Study ensures that new battery designs will comply with existing and emerging standards and regulations for safety, performance, and reliability. ||||
| Technological Innovation | Study in this stage also explores the integration of new technologies, such as solid-state electrolytes or advanced battery management systems, to improve performance and reliability.<br>Early-stage development includes creating prototypes and conducting proof-of-concept tests to validate the design and specifications. ||||
| Cross-Disciplinary Collaboration | Given the complexity of EV batteries, study at the concept and definition stage is often cross-disciplinary, involving collaboration between chemists, materials scientists, electrical engineers, and environmental scientists to develop a well-rounded battery concept. ||||

Table 3.2.2 Breakdown of the key studies of EV battery reliability lifecycle framework: Design and Development

| Character | Inherent Reliability | Stage | Design and Development |
|---|---|---|---|
| Highlights | This stage requires a multidisciplinary approach, combining insights from chemistry, materials science, mechanical engineering, electrical engineering, environmental science, and regulatory expertise. ||||
| Contents | Current research examples ||||
| Performance Optimization | Studies here focus on optimizing the design for the highest possible performance within the constraints of cost, size, weight, and intended use. This can include work on increasing energy density, improving charge rates, and enhancing thermal management systems to prevent overheating. ||||
| Advanced Materials Research | The development stage often involves researching new materials or innovative combinations of materials that can lead to better battery performance. This includes cathodes with higher lithium content, anodes made from novel materials like silicon or graphene, and electrolytes that are more stable or conductive. ||||

| | |
|---|---|
| Safety and Durability Studies | Safety is a paramount concern in battery design, leading to extensive research into features that can prevent or mitigate the effects of battery failure. <br> Durability studies ensure that the battery will retain its performance characteristics over many charging cycles and under various environmental conditions. |
| Modelling & Simulation | Before physical prototypes are created, computer models and simulations play a significant role in the design and development stage. These tools help predict how the battery will perform and identify potential issues with the design, such as areas that may experience too much heat or mechanical stress. |
| Prototype Testing | Once a design has been sufficiently modelled and simulated, prototype batteries are produced for testing. This phase includes both bench testing under controlled conditions and field testing in actual EVs to see how the battery performs in real-world scenarios. |
| Design for Manufacturability | Research also focuses on ensuring that the battery design can be efficiently and cost-effectively manufactured at scale. This can include studies on the assembly process, automation, and quality control measures. |
| Environmental Impact Assessments | At this stage, studies also assess the environmental impact of the battery design, considering the full lifecycle from raw material extraction to end-of-life disposal or recycling. This can lead to design choices that favour materials and processes with a lower environmental footprint. |
| Regulatory Compliance and Standards | Study ensures that the battery design will comply with all relevant industry standards and government regulations, which can vary widely from one region to another. Compliance is critical to ensure the battery will be legally allowed to be sold and used in its intended markets. |
| Iterative Design Improvements | The design and development phase is typically iterative, with each round of testing leading to refinements and improvements in the design. Research during this phase is geared towards continuously improving the battery design based on test results and evolving performance targets. |

Table 3.2.3 Breakdown of the key studies of EV battery reliability lifecycle framework: Manufacture and Production

| Character | Inherent Reliability | Stage | Manufacture and Production |
|---|---|---|---|
| Highlights | This stage requires a careful balance between efficiency, cost, and quality, with research playing a vital role in navigating these sometimes-competing priorities. | | |
| Contents | Current research examples | | |
| Manufacturing Process Optimization | Study in this area targets the enhancement of manufacturing processes to maximize efficiency and minimize defects. Automation of battery cell assembly lines is a focus, aiming to improve consistency and reduce the potential for human error. | | |
| Quality Control and Assurance | Quality control is pivotal to ensure that each battery adheres to specified reliability standards. Innovative quality assurance techniques are employed, leveraging real-time monitoring and predictive analytics to pre-emptively address defects. | | |
| Testing and Validation | Batteries undergo rigorous testing to confirm performance and safety standards are met. The development of accelerated life testing methods allows manufacturers to simulate long-term usage, identifying potential failure modes early on. | | |
| Material Sourcing & Supply Chain Management | Studies in this realm involve securing a stable supply of quality materials, managing the supply chain effectively, and developing strategies for risk mitigation against supply disruptions. | | |
| Environmental & Safety Standards Compliance | Amidst increasing environmental concerns, research assesses the ecological impact of manufacturing processes. Cleaner production techniques are explored to meet stringent environmental regulations. | | |

| | |
|---|---|
| Scalability & Expansion Studies | Study is conducted to address the challenges of scaling up production capacity. These studies explore how to expand rapidly while maintaining quality, particularly as EV adoption rates grow. |
| Cost Reduction Strategies | Reducing the cost of EV batteries is essential for broader market accessibility. Research into cost-effective manufacturing processes is conducted to find ways to economize without compromising on reliability. |
| Innovations in Manufacturing Technology | The adoption of new manufacturing technologies and materials that can enhance performance is a key area of focus. For example, integrating solid-state electrolytes in mass production settings is an ongoing research topic. |
| Workforce Training & Development | Ensuring that the workforce is skilled and adequately trained is crucial for high-quality manufacturing. Effective training programs are essential to equip workers with the necessary skills for producing reliable batteries. |

Table 3.2.4 Breakdown of the key studies of EV battery reliability lifecycle framework: Distribution, Installation and Commissioning

| Character | Inherent Reliability | Stage | Distribution, Installation and Commissioning |
|---|---|---|---|
| Highlights | These studies are integral to the overall success of the EV battery's operational lifecycle, helping to guarantee that the battery's inherent reliability is preserved up to the point of use. | | |
| Contents | Current research examples | | |
| Distribution Logistics | Studies in this area focus on developing optimized logistics strategies to ensure batteries are transported safely and efficiently from the manufacturing plant to the installation site. Research might explore packaging methods that protect against environmental factors like temperature extremes or physical shocks during transit. | | |
| Installation Procedures | Study here looks at the development of standardized installation procedures. Proper installation is vital to maintaining the battery's integrity and ensuring it operates as expected. Best practices for installation might involve extensive training for technicians and the use of specialized tools and equipment. | | |
| Quality Assurance Post-Distribution | Quality assurance doesn't end at the factory door; studies may investigate methods for maintaining quality control during transit and installation. This could include tracking systems that monitor the battery's condition throughout the distribution chain or post-installation testing protocols. | | |
| Impact of Storage Conditions | Batteries may be stored before installation, and research in this domain examines how various storage conditions affect their reliability. This includes studying the effects of long-term storage on battery charge state, health, and subsequent performance. | | |
| Integration with Vehicle Systems | Once installed, the battery must be integrated with the vehicle's systems. Studies here focus on ensuring that this integration goes smoothly, with research into system compatibility, software updates for battery management systems, and final quality checks before the vehicle is deemed road ready. | | |
| Customer Education | Educating the end-user about proper battery use and maintenance begins at this stage. Study into effective customer education programs can lead to the development of resources like user manuals, instructional videos, and interactive apps. | | |
| Feedback Loops for Continuous Improvement | Study also involves setting up feedback mechanisms to inform continuous improvement in the manufacturing process. This includes studying how information gathered during distribution and installation can be used to identify potential areas for enhancement in battery design and production. | | |

Table 3.2.5 Breakdown of the key studies of EV battery reliability lifecycle framework: Operation and Maintenance - 1.n *th* lifetime[2]

| Character | Use Reliability | Stage | Operation and Maintenance - 1.n *th* lifetime |
|---|---|---|---|
| Highlights | Study in this phase is essential to ensure that EV batteries can reliably meet the expectations of consumers and the demands of daily operation. | | |
| Contents | Current research examples | | |
| Integration of Inherent and Use Reliability | Studies focus on how inherent reliability (defined during the design and manufacturing phases) and use reliability (observed during actual operation in vehicles) interact to determine the overall reliability of the battery. This involves assessing how design choices and manufacturing quality impact long-term performance and how real-world usage conditions influence the degradation and failure rates of batteries. | | |
| Long-Term Performance Monitoring | Research in this area involves the continuous monitoring of battery performance over its operational life. This includes tracking changes in capacity, power output, efficiency, and other key performance indicators over time, under varying operational conditions. | | |
| Predictive Analytics for Reliability | Advanced data analytics and machine learning models are employed to predict future reliability based on historical performance data. This predictive approach helps in proactive maintenance and timely intervention to extend battery life. Some typical studies are focusing on SOC, SOH, etc. | | |
| Real-World Performance Metrics | Studies measure how the battery performs under various driving conditions, such as urban stop-and-go traffic, highway speeds, and under different climate conditions. This includes tracking metrics like range per charge, power output, and state of health over time. | | |
| User Behaviour & Usage Patterns | Study investigates the impact of user behaviour on battery reliability. Different charging habits, such as frequency of charges, depth of discharge, and reliance on fast charging, can significantly affect battery life and performance. | | |
| Battery Management Systems (BMS) | Advanced BMS are crucial for maintaining battery health. Studies focus on improving BMS algorithms to optimize charging and discharging processes, thermal management, and energy efficiency to extend the battery's useful life. | | |
| Maintenance Practices | The effect of maintenance routines on battery reliability is another area of study. Regular health checks, software updates, and condition-based maintenance can prolong the battery's life and ensure consistent performance. | | |
| Degradation Analysis | Research into battery degradation seeks to understand and model how and why batteries lose capacity and power over time. This includes studying the physical and chemical changes within battery cells and developing strategies to minimize degradation. | | |
| Failure Mode Analysis | Identifying and analysing potential failure modes helps improve use reliability. Studies might include stress testing and post-mortem analysis of failed batteries to better understand the causes of failure and to improve future designs. | | |
| Data Analytics & Predictive Maintenance | With the advent of big data, studies use analytics to predict battery life and recommend maintenance. Machine learning models can forecast when a battery might fail or require service, allowing for proactive maintenance. | | |
| Impact of External Factors | External factors such as extreme temperatures, humidity, and mechanical vibrations can affect battery reliability. Research aims to characterize these impacts and develop batteries that are more resilient to such factors. It is crucial for developing batteries that are robust and reliable across a wide range of operating environments. | | |
| Sustainability and Second-Life Applications | As batteries approach the end of their use in vehicles, research shifts to potential second-life applications. Studies determine the viability of using degraded batteries for less demanding applications, like stationary energy storage, and assess the reliability in these new roles. | | |
| User Feedback & Warranty Claims | Customer feedback and warranty claim data are valuable sources of information on battery reliability. Analysing this data helps identify common issues and areas for improvement. This feedback loop is essential for continuous improvement in battery design and manufacturing processes. | | |

---

[2] Some contents may be also included in "Repurpose" stage.

| End-of-Life Management | Study in this area includes studies on the efficiency and effectiveness of recycling processes and the potential for second-life applications. |
|---|---|

Table 3.2.6 Breakdown of the key studies of EV battery reliability lifecycle framework: Operation and Maintenance - 2nd lifetime (Repurpose)

| Character | Use Reliability | Stage | Operation and Maintenance - 2nd lifetime (Repurpose) |
|---|---|---|---|
| Highlights | This field addresses the potential for extending the useful life of EV batteries beyond their initial automotive application, delving into various aspects of their performance, adaptability, and overall viability in new roles. ||||
| Contents | Current research examples ||||
| Assessment of Battery Health for Repurposing | Studies in this area focus on evaluating the remaining capacity and performance metrics of used EV batteries. This involves determining how much of the original capacity and power the batteries retain and whether they are suitable for less demanding applications compared to their initial automotive use. ||||
| Optimization for Second-Life Applications | study is directed towards optimizing the performance of repurposed batteries for specific second-life applications, such as stationary energy storage systems for homes, businesses, or utility-scale projects. This includes modifying battery management systems and adapting the batteries to new charging and discharging patterns different from those in EVs. ||||
| Safety and Reliability in New Roles | Given the altered operational context, studies investigate the safety and reliability of these batteries when used for energy storage or other purposes. This involves understanding the risks of battery failure modes in these new environments and implementing appropriate safety measures. ||||
| Economic and Environmental Analysis | Economic feasibility studies assess the cost-effectiveness of repurposing batteries compared to manufacturing new energy storage systems. Environmental impact studies evaluate the benefits of extending the life of batteries, such as reducing waste and the demand for new raw materials. ||||
| Integration with Renewable Energy Systems | A significant area of interest is how repurposed EV batteries can be integrated with renewable energy sources like solar and wind power. Study in this domain explores the synergies between intermittent renewable energy production and the storage capabilities of these batteries. ||||
| Standardization & Regulatory Compliance | Studies also examine the need for standardization in the process of repurposing batteries, ensuring compatibility with existing energy systems and compliance with electrical and safety standards. ||||
| Development of Second-Life Market | Study includes the exploration of business models and market opportunities for second-life batteries. This encompasses the logistics of collection, testing, refurbishment, and distribution of repurposed batteries. ||||
| Lifecycle Assessment | Lifecycle assessments are crucial in this stage to understand the full environmental impact of repurposing batteries, from collection and transportation to eventual disposal after their second life. ||||

Table 3.2.7 Breakdown of the key studies of EV battery reliability lifecycle framework: Recycle

| Character | Use Reliability | Stage | Recycle |
|---|---|---|---|
| Highlights | The recycle stage studies encompass a wide range of interdisciplinary research areas, all aiming to improve the sustainability, efficiency, and profitability of EV battery recycling, while ensuring environmental safety and compliance with evolving regulations. | | |
| Contents | Current research examples | | |
| Material Recovery and Recycling Processes | **Efficiency of Material Recovery:** Research is conducted on the efficiency of recovering valuable materials such as lithium, cobalt, nickel, and manganese from used EV batteries. Studies aim to improve the yield and purity of recovered materials.<br>**Advanced Recycling Technologies:** There are continuous efforts to develop new technologies that can more efficiently disassemble, and process spent batteries, such as hydrometallurgical, pyrometallurgical, and bio-hydrometallurgical techniques.<br>**Design for Recycling:** Studies look at how the initial design of batteries can be optimized to facilitate easier and more complete recycling at the end of their useful life. | | |
| Environmental Impact of Recycling | **Life Cycle Assessment (LCA):** Researchers use LCA to evaluate the environmental impacts associated with the recycling of EV batteries, including energy consumption, greenhouse gas emissions, and the potential release of toxic substances.<br>**Regulatory Compliance:** With a growing emphasis on environmental regulations, studies also review compliance with international standards and regulations governing the disposal and recycling of batteries | | |
| Economic and Policy Considerations | **Cost-Benefit Analysis:** There is research on the economic aspects of recycling, including the costs of recycling processes compared to the value of recovered materials.<br>**Policy Incentives:** Policy-driven research examines the effectiveness of incentives, such as extended producer responsibility (EPR), in promoting the development of efficient recycling ecosystems. | | |
| Second-Life Applications | **Feasibility Studies:** Research is done to determine the feasibility of using recycled batteries in second-life applications, such as stationary energy storage systems, and their reliability in these new roles.<br>**Performance Metrics:** Studies measure the performance of repurposed batteries, assessing factors like capacity fade, energy efficiency, and overall reliability after recycling. | | |
| Innovation and Future Directions | **Cutting-edge Techniques:** Innovative recycling methods, such as direct recycling that aims to preserve the electrochemical properties of battery materials, are a focus of current research.<br>**Scalability and Infrastructure:** As the number of EVs on the road increases, research also looks at how recycling infrastructure can be scaled up to handle larger volumes of spent batteries. | | |

# Chapter 4. System Cognition of EV Battery Reliability: Integrating Reliability Ecosystem and Lifecycle Framework

This chapter explores a novel approach to understanding the reliability system of EV batteries. Such "cognition" of reliability system consists of an integration of the reliability ecosystem and lifecycle frameworks for (EV battery's) system reliability. Therefore, it begins by delineating the distinct characteristics of a reliability system and system reliability. Then it presents a multidimensional system cognition of EV battery reliability that correlates the fundamental elements of point, line, surface, system, and system of systems (SoS). A key focus of this chapter is the conceptualization of the EV battery system reliability lifecycle as a hyperplane projection of its reliability system, where time serves as a critical dimension. This viewpoint allows for a dynamic understanding of reliability system, considering how it evolves and is influenced over the battery's lifespan. This comprehensive approach not only fosters better understanding about a holistic system cognition of EV battery reliability constructed but also guides strategic decision-making in the realm of EV battery system optimization (RQ3).

## 4.1 Reliability System and System Reliability of EV Batteries

The concept of a "reliability system" in the context of EV batteries encompasses a comprehensive array of methods, processes, tools, and practices dedicated to ensuring and enhancing the reliability of these batteries. This system includes various elements such as reliability engineering techniques, maintenance strategies, quality control measures, testing procedures, and data analysis methodologies. The emphasis here lies in the systematic approach towards achieving, sustaining, and improving reliability. It involves creating an infrastructure and methodology focused on ensuring that EV batteries perform reliably over their intended lifespan. Essentially, a reliability system addresses the "how" of reliability – the implementation of specific practices and principles to attain reliable outcomes.

Conversely, "system reliability" in the realm of EV batteries pertains to the probability or likelihood of these batteries performing their required functions without failure over a specified period under predefined conditions. It is a metric or characteristic of the system, typically measured through indicators like Mean Time Between Failures (MTBF) or failure rates. System reliability zeroes in on the actual performance and dependability of the EV batteries, representing the tangible outcome or attribute experienced by users. This term focuses on the "what" of reliability – the actual performance and reliability of the battery system as observed in real-world applications.

In essence, while a "reliability system" in EV batteries is about the strategic framework and methodologies employed to ensure reliability, "system reliability" refers to the demonstrable reliability performance of the battery system. The former encompasses a proactive and strategic approach, utilizing a variety of tactics and techniques, whereas the latter signifies a quantifiable measure or characteristic that reflects the battery system's likelihood to operate reliably.

## 4.2 Integrating Reliability Ecosystem and Lifecycle frameworks: Cognation from Point to System of Systems (SoS)

This chapter presents an integrated view of the reliability ecosystem and lifecycle frameworks of EV batteries, conceptualized through the geometry of a triangular prism, as depicted in Fig 4.1. This geometric representation serves as a simplified model for understanding the complex structure and dynamics of reliability within EV battery systems.

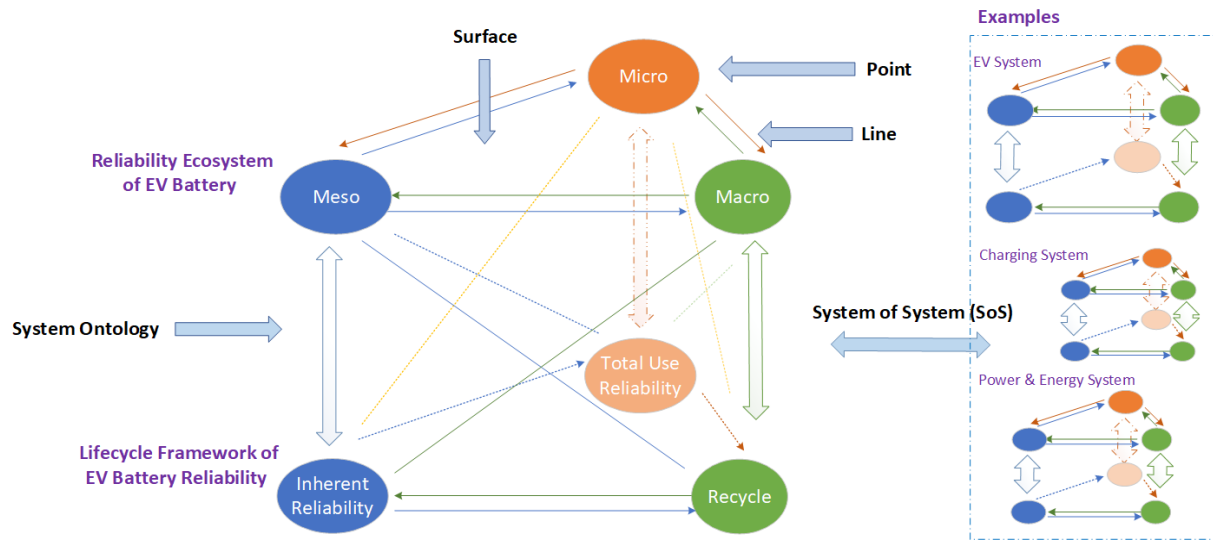

Fig 4.1 The Reliability System of EV Batteries: from point to System of Systems (SoS)

Assuming a stationary status, the triangular prism embodies the multidimensional structure of the reliability system. It encompasses points, lines, surfaces, and the body—each representing a different aspect of the reliability system and lifecycle.

- **Points - The Fundamentals of Reliability:** Points are depicted as the basic building blocks of the reliability system. Within the context of EV batteries, these points represent entities across the Micro, Meso, and Macro perspectives outlined in Chapter 2. They include individuals or components at the micro-level, industrial entities at the meso-level, and societal elements at the macro-level. From the lifecycle framework discussed in Chapter 3, points also symbolize crucial attributes or processes such as Inherent Reliability, Total Use Reliability, or the recycling processes at the end of a battery's life. These points are the foundation upon which the reliability system is built.
- **Lines - The Connections that Bind:** Lines within the triangular prism model define the relationships among points. They represent the interactions between entities, attributes, and processes, or the constraints that govern the system's functioning and reliability. These lines are pivotal as they form the connections that allow for the flow of energy, information, and materials between points, facilitating the system's integrated functioning.
- **Surfaces - The Networks and Interactions:** The prism consists of six surfaces, each symbolizing a network of points and lines—a subsystem within the greater reliability ecosystem. These surfaces can be seen as the interactions between various components and processes within the EV battery's lifecycle, including design, manufacturing, operation, and recycling. They also represent the interplay between different perspectives within the ecosystem, where micro-level interactions build up to meso and macro-level dynamics.
- **Body - The Reliability System Ontology:** The body of the triangular prism is representative of the entire reliability system of EV batteries. It encompasses the aggregated knowledge and interactions encapsulated within the points, lines, and surfaces. The body is a holistic embodiment of the system's reliability, illustrating how components and processes coalesce to form a complete, functioning system.
- **System of Systems (SoS) - Beyond Individual Systems:** When these systems interact with other systems—engaging in collaborative complexity—they form a System of Systems (SoS). As Shown in Fig 4.1, some examples of the other systems include: EV system, Smart Grid system, Charger Station system, etc. The SoS cognition involves understanding how each system, with its unique goals and metrics, contributes to a larger, more complex goal when integrated with other systems. This is where emergent behaviours and functionalities arise, providing new capabilities that are not possible within individual systems.

- **From Components to Cognition:** Each point within the system has its goals, leading to metrics for quantifiable measurement. Lines can evolve into platforms facilitating interactions, while surfaces may give rise to industry segments or research domains. This chapter elucidates how each component of the reliability system, from the micro-level points to the macro-level SoS, is integrated into a cohesive framework. It explores the implications of this integration for advancing the reliability, performance, and sustainability of EV batteries.

This chapter weaves together the reliability ecosystem and lifecycle frameworks into a comprehensive understanding of EV battery reliability. It lays out a structured model that captures the essence of a reliability system and demonstrates how it operates within the complex interrelations of a System of Systems.

## 4.3 Hyperplane Projection: Insights on the Intrinsic of Reliability System Optimization of EV Batteries

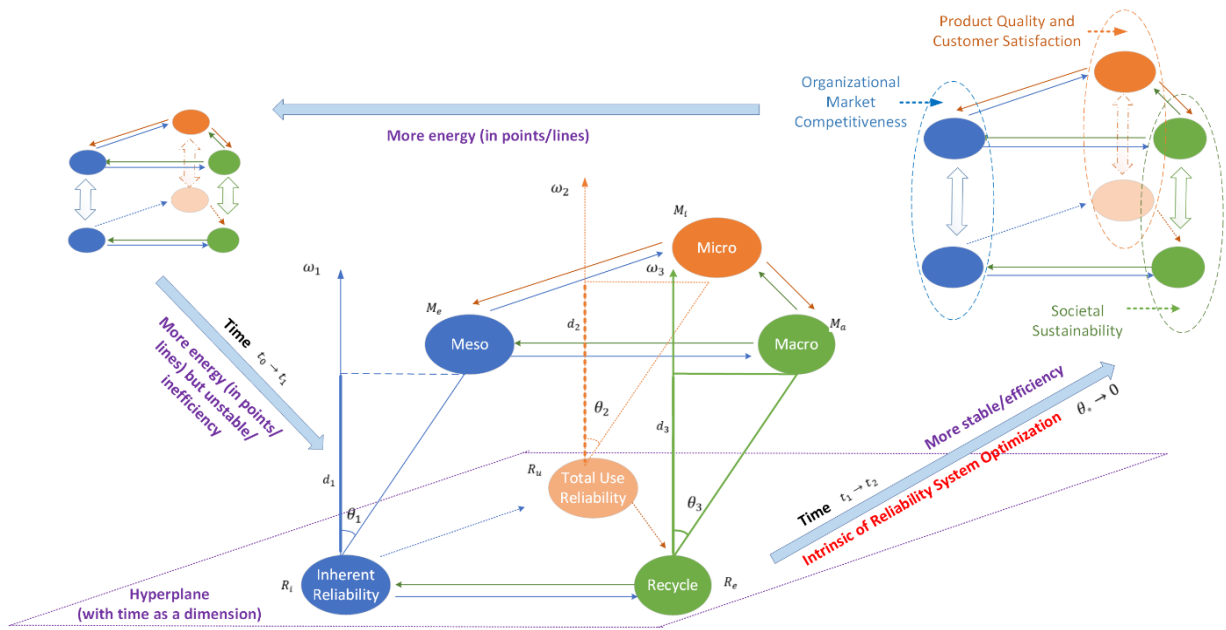

Fig 4.2 Hyperplane Projection and the Intrinsic of Optimization

Section 4.2 established a stationary framework for the reliability system of EV batteries, characterized by a structured interplay between points, lines, and surfaces. However, this depiction did not account for the critical dimension of time. This section explores the concept of time as a pivotal factor that transforms the stationary status into a dynamic one, revealing the intrinsic nature of reliability system optimization (Fig 4.2).

- **Lifecycle Framework as a Hyperplane:** The lifecycle framework can be conceptualized as a hyperplane—a multidimensional space that captures the complex interactions within the reliability system. On this hyperplane, key elements can be projected such as Inherent Reliability ($R_i$), Total Use Reliability ($R_u$), and Recycling ($R_e$). These projections represent the influences of the points of Micro ($M_i$), Meso ($M_e$), and Macro ($M_a$) levels within the reliability ecosystem. $\boldsymbol{\omega}$ represents the Normal Vector; $\theta_1$, $\theta_2$, $\theta_3$ represents angles, and $d_1$, $d_2$, $d_3$ represents distance.

- **Stationary Status and Its Significance:** When the vectors representing these projections align (i.e., the angle between them is zero), a stationary status can be achieved. This alignment signifies that the reliability system is optimally meeting the needs of all stakeholders in the ecosystem. In such a state, there is harmony between what is anticipated by the ecosystem (the stakeholders' needs) and what is delivered by the reliability system (the performance of the EV batteries).
- **Strategic Implications of Projections:** The projection of the Micro level influences critical aspects such as product quality and customer satisfaction. It underscores the direct impact on end-users, emphasizing the importance of reliability in the consumer experience. The projection of the Meso level shapes organizational competitiveness. It encompasses the strategic manoeuvres of businesses within the industry, reflecting how reliability affects operational excellence and market positioning. The projection of the Macro level extends to societal sustainability. It captures the broader implications of EV battery reliability on environmental stewardship, resource management, and sustainable development.
- **Time as a Dynamic Element:** Introducing time as a dimension allows us to observe the evolution of the reliability system from a static to a nonstationary status. For instance, in Fig. 4.2, from $t_0 \to t_1$, this dynamic change is driven by the energy changes within points (such as new technological advancements or wear and tear of components) or shifts along lines (like changes in market demands or regulatory updates); from $t_1 \to t_2$, the system is more stable and efficient after system optimization. It should be noticed that, when the system is stable, the angle equals to 0. From this point, it indicates that the goal of reliability system optimization is to make $\theta_* \to 0$.
- **Energy Changes and System Optimization:** The energy changes could be literal, in terms of the actual energy capacity and performance of the batteries over time, or metaphorical, representing shifts in industry trends, technological innovation, and policy reforms. These changes necessitate continuous optimization of the reliability system to adapt and respond to new conditions, ensuring that the system maintains its effectiveness and efficiency over time.

In summary, this section provides a deep dive into the intrinsic characteristics of reliability system optimization for EV batteries (to make $\theta_* \to 0$), considering the dynamic nature of time (from $t_1 \to t_2$). It elucidates how projections on the hyperplane of the lifecycle framework can guide strategic decisions at various ecosystem levels and how the concept of time necessitates a flexible and adaptive approach to maintain optimal reliability. This section enhances our understanding of the evolving nature of EV battery reliability, offering insights that inform both current practices and future innovations.

**Chapter 5. Discussions**

Following the above critical thinking from comprehensive ecosystem perspectives, lifecycle innovation, system cognation on EV Battery reliability, Chapter 5 engages in examining the inconsistencies between theoretical and actual battery performance (Section 5.1); proposing the focus on enhancing battery reliability by emphasizing the importance of accurate sensor data through thorough assessment and calibration (Section 5.2); discussing the optimization of EV battery reliability system as a whole with global sustainability initiatives, including sustainable development goals (SDGs), environmental, social, and governance (ESG) criteria, and the "battery passport"(Section 5.3); and offering strategic insights into enhancing system reliability through continuously improvement of reliability system by proposing a Social-Industrial Large Knowledge Model (S-ILKM) framework (Section 5.4). The aim is to provide a holistic perspective for advancing EV battery reliability, aligning with the larger goals of sustainable and efficient transportation.

## 5.1 Enhancing EV Battery Reliability through examining inconsistencies: Dissecting the Gaps between Theoretical vs. Actual Reliability

It is observed that there is a considerable gap between the expected (theoretical) and the observed (actual) reliability of EV batteries. For instance, while the inherent reliability of a battery product may predict a lifespan of 8-20 years[3], its actual performance in an EV often falls short, lasting only between 5-8 years. This inconsistency is not just a matter of statistical variance but a serious concern that necessitates a deeper investigation. The significant inconsistency between the theoretical and actual reliability of electric vehicle (EV) batteries leads to a spectrum of negative consequences, as depicted in Fig 5.1. These implications range from immediate safety issues to long-term financial and environmental burdens.

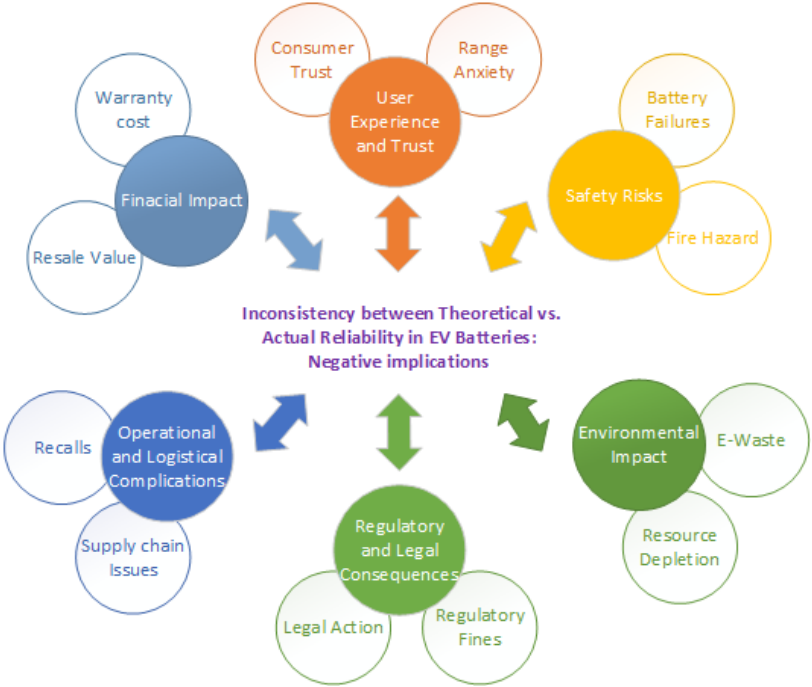

Fig 5.1 Impacts of Reliability Inconsistency in EV Batteries

Regarding safety risks, unexpected battery failures can leave drivers stranded, jeopardizing their safety. Additionally, a mismatch between the battery's actual reliability and theoretical safety models can increase the risk of overheating, potentially leading to fires. Financially, a high rate of battery failure escalates warranty claims, imposing financial strain on manufacturers. This issue also affects the resale value of EVs, as reduced reliability can diminish the market value, impacting both manufacturers and vehicle owners. From an environmental perspective, premature battery failure leads to increased electronic waste (E-Waste), as recycling or repurposing becomes less feasible. Moreover, the need for frequent replacements of short-lived batteries accelerates the depletion of raw materials. Operationally, significant reliability discrepancies may trigger product recalls, incurring substantial costs and potentially damaging the brand's reputation. Additionally, unreliability issues can disrupt the supply chain, forcing a search for alternative suppliers or technologies. In terms of user experience and trust, frequent failures or performance issues can erode consumer trust in a brand, negatively affecting future sales and market share. One common concern among users is "range anxiety" – the fear that their vehicle will run out of power before reaching its destination, especially if the batteries do not last as long as expected. Finally, regulatory and legal consequences are also a concern. Failure to meet reliability standards can lead to regulatory actions, including fines or prohibitions on selling certain products. Moreover, manufacturers may face legal action if they were aware of the reliability gap but did not take steps to resolve it.

---

[3] The lifetime of batteries is normally defined with times of charging and discharging cycles; But the end users feel in the way like "kilometres" "years".

The inconsistency between the theoretical and actual reliability of EV batteries can be attributed to a range of factors throughout their reliability lifecycle. One of the primary reasons is inherent reliability issues that arise even before the battery is operated in vehicles. This includes challenges in battery component sourcing, where the quality and consistency of materials can vary significantly. Manufacturing variations, both in the production of the battery itself and the vehicle manufacture, also contribute to these discrepancies. For instance, two batches of batteries with slight differences in material composition or assembly can exhibit varied performance and longevity. Before installation, the way batteries are stored, and the measurement capabilities of sensors used to monitor them play a crucial role. Batteries stored in conditions not conducive to their longevity or monitored with less accurate sensors might not retain their intended quality. Once the batteries are in use, their reliability is further challenged by the complexity of real-world scenarios compared to controlled lab conditions. Environmental factors like extreme temperatures or user behaviours such as frequent rapid charging and discharging can accelerate battery degradation, deviating from theoretical predictions. Moreover, the integration of the battery system within the vehicle, including its interaction with other systems and the battery management software, can introduce complexities not accounted for in theoretical models. As the battery ages, wear and tear, along with chemical degradation, further reduce its performance, often at a rate faster than anticipated. Economic and business factors, such as cost constraints and market pressures, can also influence the degree to which batteries are optimized for long-term reliability. One key insight for enhancing battery system reliability is the importance of integrating real-world usage data and feedback into the battery design and manufacturing process. For example, if data shows that batteries degrade faster under certain environmental conditions, manufacturers can adjust the battery composition or protective measures to mitigate this effect. Similarly, understanding user behaviour patterns can lead to designing batteries that are more resilient to specific usage patterns. This approach of continuous learning and adaptation, informed by actual usage data, can significantly close the gap between theoretical and actual reliability.

## 5.2 Optimizing EV Battery Reliability through Evaluating and Calibrating the Measurement Capabilities of EV Battery Sensors

The measurement capabilities of sensors within electric vehicle (EV) battery systems are a critical, yet often overlooked, aspect of ensuring overall system reliability. These sensors play a vital role in monitoring various parameters such as temperature, voltage, current, and state of charge (SOC), which are essential for accurate battery management. Over time, the precision and accuracy of these sensors can degrade due to factors like environmental conditions, aging, or mechanical stress. This degradation can lead to unreliable data being fed into the battery management system (BMS), resulting in inaccurate diagnoses and prognoses of the battery's health and performance.

The need for assessment and calibration of these sensors cannot be overstated. Proper calibration ensures that sensor readings remain accurate and consistent over time. This process involves adjusting the sensor output or the data interpretation methods to align with standard or reference measurements. Without such evaluation and calibration, there is a risk that the system might make decisions based on faulty data, which can lead to a range of issues—from inefficient battery utilization to safety hazards. For instance, if a temperature sensor starts to show readings that are higher or lower than the actual battery temperature, the BMS might unnecessarily trigger cooling systems or fail to prevent overheating, both of which can affect battery life and safety. Similarly, inaccurate SOC readings might lead to undercharging or overcharging of the battery, reducing its lifespan and efficiency.

The consequences of not evaluating and calibrating sensors can be significant. Incorrect diagnostic and prognostic results can lead to premature battery failures, reduced operational efficiency, increased maintenance costs, and even safety risks. Therefore, it is crucial to integrate sensor evaluation and calibration into the operation and maintenance stage of EV battery systems during the lifecycle management. This practice not only ensures the accuracy and reliability of sensor data but also contributes to the overall health and longevity of the battery system, ultimately enhancing the reliability and safety of electric vehicles.

### 5.3 Optimizing EV Battery Reliability System as a whole: Aligning with SDGs, ESG, and the "Battery Passport" Framework

The reliability of electric vehicle (EV) batteries holds a significant place in the broader context of sustainable development goals (SDGs), environmental, social, and governance (ESG) criteria, and the emerging concept of the "battery passport". EV battery reliability is not an isolated technical feature; the integration of EV battery reliability within these frameworks demonstrates the multi-dimensional impact of technological advancements in achieving global sustainability goals.

When considering the novel EV battery reliability ecosystem, the framework of EV battery reliability lifecycle, and the holistic system cognition for reliability system optimization in the context of SDGs, ESG criteria, and the "battery passport", it's imperative to acknowledge the unique nature of EV batteries as opposed to general assets. This is due to several reasons:

- **Complexity and Interconnectivity at Micro, Meso, and Macro Levels:** The EV battery reliability ecosystem encompasses a wide range of stakeholders and factors at different levels, each playing a crucial role in the overall reliability and performance of the batteries. This includes individual consumers (micro-level), manufacturers and industry players (meso-level), and broader societal and environmental impacts (macro-level). Addressing SDGs, ESG criteria, and implementing a "battery passport" system requires a deep understanding of these interconnections and the cumulative impact they have on the sustainability and ethical implications of EV batteries.
- **Lifecycle Framework Inclusivity:** Traditional asset frameworks often overlook stages like "Zero"-Life reliability, reuse, repurpose, and recycle. However, these stages are vital in the context of EV batteries for achieving sustainability goals. The comprehensive lifecycle approach of EV batteries ensures that every stage, from production to end-of-life management, aligns with sustainability principles, thereby supporting various SDGs and ESG objectives. This includes reducing waste, promoting resource efficiency, and ensuring responsible production and consumption patterns.
- **Holistic System Cognition for Reliability System Optimization:** The holistic approach to system cognition in EV battery reliability acknowledges the dynamic nature of these systems and their evolution over time. This approach is crucial for effective reliability system optimization, ensuring that solutions are adaptable, resilient, and sustainable in the long term. It aligns with the principles of the "battery passport", which seeks to track and optimize the battery's lifecycle for maximum efficiency and minimum environmental impact.
- **Alignment with Global Sustainability Goals:** The novel ecosystem and lifecycle framework of EV batteries are directly aligned with several SDGs, including clean energy, sustainable cities, responsible consumption, and climate action. Enhancing reliability through this comprehensive approach ensures that EV batteries contribute effectively to these goals.

- **Compliance with ESG Criteria:** ESG criteria emphasize responsible and sustainable business practices. A holistic approach to EV battery reliability, considering the entire ecosystem and lifecycle, ensures that these criteria are met not just in isolated instances but as an integral part of the business model. This includes ethical sourcing of materials, ensuring labour welfare, minimizing environmental impact, and demonstrating strong corporate governance.
- **Advancing Circular Economy Principles:** The novel lifecycle approach, particularly in stages like reuse and recycling, is crucial for advancing circular economy principles. This approach is essential for reducing the ecological footprint of EV batteries, thereby aligning with the goals of sustainability and responsible resource management.

In summary, considering the novel EV battery reliability ecosystem, lifecycle framework, and holistic system cognition is crucial when addressing SDGs, ESG, and "battery passport" due to the complex, interconnected nature of EV batteries and their significant impact on sustainability and ethical practices. This approach ensures that optimization in battery reliability system are not just technical improvements but contribute to broader global goals of sustainability and responsible innovation.

### 5.4 Advancing to a Social-Industrial Large Knowledge Model (S-ILKM) Framework

The transition from traditional Industrial Large Knowledge Models (ILKMs) (Lee & Su, 2023) to an expansive Social-Industrial Large Knowledge Model (S-ILKM) marks a significant evolution in the application of large language models (LLMs) within Industry 4.0 and smart manufacturing (Fig 5.2). ILKMs focus on integrating domain-specific knowledge into LLMs to address complex challenges in industrial settings, particularly in smart manufacturing. However, these models predominantly cater to industrial needs, with a concentration on meso-level inputs and outputs.

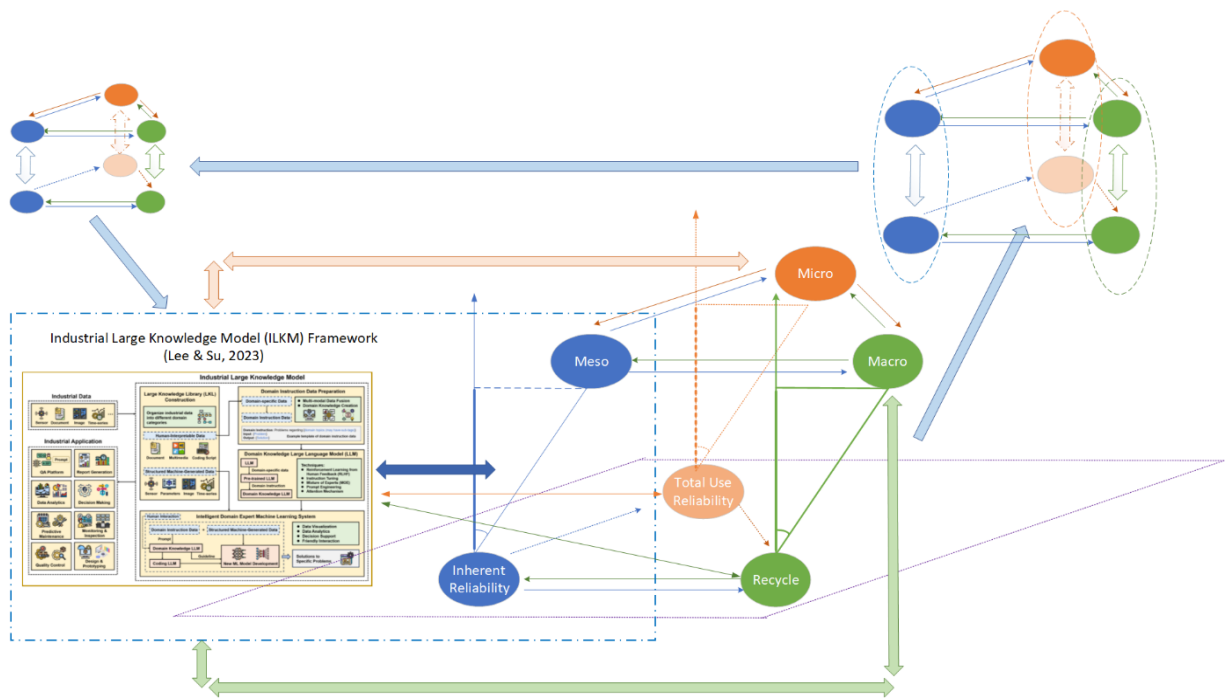

Fig 5.2 An expansive Social-Industrial Large Knowledge Model (S-ILKM)

In the context of EV battery reliability and drawing insights from the reliability system cognition explored in Chapter 4, it becomes evident that ILKM's focus is somewhat could be extended to a broader scope. While they undoubtedly make manufacturing systems smarter and enhance inherent reliability at the meso level, there's a growing need for a more encompassing approach. This need is highlighted by the dynamic energy changes at the micro and macro levels that occur over time, indicating that a continuous improvement in the reliability system is essential (Fig 5.2).

S-ILKM extends beyond ILKM's industrial focus, integrating insights from both micro and macro levels. This approach broadens the data scope to include both industry-specific, private data (like quality and reliability data) and diverse, public open-source data (such as market trends, regulations, and user feedback). The purpose of S-ILKM is to improve systems holistically, not just for industrial tasks or language-related challenges. This model fuses in-depth domain knowledge specific to EV battery industries with broader societal and global contexts.

Other key aspects of S-ILKM include:

- **Data Diversity:** Combining private, domain-specific data with public, diverse textual data, S-ILKM offers a more comprehensive view for system improvement.
- **Expanded Purpose:** S-ILKM is designed for broader system improvements, transcending the specific industrial tasks of ILKM and the language-centric focus of LLMs.
- **Knowledge Fusion:** It integrates detailed domain knowledge pertinent to EV battery industries with broader individual experiences and societal insights.
- **Ecosystem Integration:** Unlike ILKM, S-ILKM aligns with the entire ecosystem, encompassing micro, meso, and macro levels for a more holistic approach.
- **Real-Time Decision Making:** S-ILKM is equipped for immediate decision-making within the battery reliability system, expanding its utility beyond just industrial applications.

In summary, the S-ILKM framework represents a leap forward in knowledge modelling, offering a comprehensive, integrated approach that considers the intricate interplay of individual, industrial, and societal factors in optimizing EV battery reliability. This advanced model is pivotal for addressing the dynamic complexities of modern EV battery systems and aligns well with the evolving landscape of sustainable transportation technology.

**Chapter 6. Conclusion**

The study ventured beyond traditional approaches to battery reliability, advocating for a holistic, multi-faceted view that interweaves the reliability ecosystem, lifecycle innovation, and system cognition into a cohesive narrative. In Chapter 2, it explored the crucial quest for high reliability in EV batteries, dissecting the complex reliability ecosystem across micro, meso, and macro perspectives. This comprehensive analysis revealed the impact of battery reliability, extending from individual consumer satisfaction to broader industrial competitiveness and environmental sustainability. Chapter 3 delved into the operational reliability of EV batteries, transitioning from general asset frameworks to a more nuanced EV battery lifecycle. The introduction of the novel "Zero"-Life reliability stage and the emphasis on use and reuse, repurpose, and recycle stages provided a fresh lens through which to view battery reliability, enriching our understanding and approach to lifecycle management. The study's innovative approach continued in Chapter 4, where it integrated the reliability ecosystem with the lifecycle framework to form a dynamic system cognition of EV battery reliability. This multidimensional perspective, visualized through the metaphor of a triangular prism, facilitated a deeper understanding of the intrinsic nature of reliability system optimization. In Chapter 5, its discussions synthesized these insights, addressing the inconsistencies between theoretical predictions and actual performance, emphasizing the critical role of evaluating and optimizing the measurement capabilities of the sensors involved, aligning EV battery reliability with global sustainability initiatives (including SDGs, ESG criteria, and the "battery passport" framework), and proposing the transformative Social-Industrial Large Knowledge Model (S-ILKM) framework.

As this study concludes, it becomes evident that the enhancement of EV battery reliability requires a shift from isolated, component-focused approaches to a holistic system optimization strategy. This broader perspective not only addresses the immediate technical challenges but also aligns with global sustainability objectives. The study's insights and recommendations pave the way for future research and practical applications in sustainable transportation and EV technology, emphasizing the critical role of EV battery reliability in shaping a sustainable, efficient, and environmentally friendly future. In conclusion, this study contributes to the ongoing discourse in EV technology by offering a comprehensive, innovative framework for understanding and improving EV battery reliability. By embracing a holistic view that integrates ecosystem analysis, lifecycle innovation, and system cognition, this study sets a new benchmark in the pursuit of sustainable, high-performing, and durable EV batteries, crucial for the advancement of sustainable transportation worldwide.